\documentclass[aps,prl,twocolumn,superscriptaddress,showpacs]{revtex4-2}

\usepackage{amsmath,amssymb}
\usepackage{graphicx}
\usepackage{bm}
\usepackage[colorlinks=true,linkcolor=blue,citecolor=blue,urlcolor=blue]{hyperref}

\begin{document}

\title{A Scalable Configuration-Interaction Impurity Solver via Active Learning}

\author{Jeongmoo Lee}
\author{Ara Go}
\email{arago@jnu.ac.kr}
\affiliation{Department of Physics, Chonnam National University, Gwangju, Korea}

\begin{abstract}
Finite-Hamiltonian impurity solvers provide direct real-frequency spectra and a natural route to enlarged impurity Hamiltonians, but their applicability is limited by the rapid Hilbert-space growth with the number of bath or other added one-particle orbitals.
We introduce active-learning adaptive-truncation configuration interaction (AL-ATCI), an extension of adaptive-truncation configuration interaction (ATCI), that identifies the determinant manifold relevant to the correlated state.
The approximation is systematically controlled by the size of the selected configuration subset, which also provides an internal convergence parameter when no external benchmark is available.
Over the benchmark range studied here, the computational cost grows only weakly with bath size, because enlarging the bath mainly expands the combinatorial determinant space rather than the physically relevant manifold.
In dynamical mean-field-theory benchmarks for the one-dimensional Hubbard model, AL-ATCI reproduces exact-diagonalization accuracy and extends cellular calculations to ten correlated impurity orbitals, corresponding to ten cluster sites in this single-orbital case.
For a three-orbital rotationally invariant Sr$_2$RuO$_4$ impurity problem, we demonstrate systematic convergence of dynamical quantities and a highly compressed configuration space as the number of bath orbitals is increased from 9 to 18.
These results substantially alleviate the bath-discretization bottleneck of exact-diagonalization- and configuration-interaction-based impurity solvers and make large-bath and enlarged-orbital calculations more practical.
\end{abstract}

\maketitle

% Introduction =====================================================================================
\textit{Introduction.}---Dynamical mean-field theory (DMFT) is a central framework for treating strong local electronic correlations in quantum materials, mapping a lattice problem onto a quantum impurity model coupled to a self-consistent bath~\cite{Georges1996,Kotliar2006}. 
The accuracy and practicality of DMFT therefore hinge on the impurity solver. 
Continuous-time quantum Monte Carlo solvers~\cite{Werner2006,Gull2011} are powerful and continue to advance for general multiorbital interactions~\cite{Eidelstein2020} and nonequilibrium transport~\cite{Erpenbeck2024}, while numerical renormalization group methods~\cite{Wilson1975,Bulla2008} provide excellent low-energy resolution for problems such as Hund metals~\cite{Stadler2015,Kugler2020} and heavy-fermion quantum criticality~\cite{Gleis2025}. 
Here we focus on the complementary class of finite-Hamiltonian impurity solvers based on exact diagonalization (ED)~\cite{Caffarel1994,Capone2007,Lu2017} and configuration interaction (CI)~\cite{Zgid2011,Zgid2012,Lin2013,Lu2014,AraGo2015,AraGo2017,MejutoZaera2019,Kitatani2023}, which are sign-problem free, work directly on the real-frequency axis at zero temperature, and provide explicit Anderson impurity Hamiltonians.
Throughout, $N_c$ denotes the number of correlated impurity orbitals, excluding spin; for a single-orbital lattice model this is identical to the number of cluster sites. 
Similarly, $N_b$ denotes the total number of bath orbitals.

The explicit Hamiltonian formulation is both a limitation and an opportunity. 
Its limitation is the finite-bath approximation: the continuous hybridization function must be represented by a finite number of bath orbitals, making convergence with bath size $N_b$ essential. 
Its opportunity is that additional ligand, core, or auxiliary orbitals can be incorporated within the same Hamiltonian framework, enabling calculations of observables such as x-ray absorption spectroscopy (XAS) and resonant inelastic x-ray scattering (RIXS)~\cite{Wang2019,Hariki2020,Higashi2021}. 
In conventional ED, however, the exponential growth of the Hilbert space makes both large-bath calculations and enlarged-orbital impurity models rapidly prohibitive.

CI-based truncation and selection methods reduce this cost by working in a restricted determinant space~\cite{Coe2018,Jeong2021,Zgid2011,Zgid2012,Lin2013,Lu2014,AraGo2015,AraGo2017,MejutoZaera2019,Kitatani2023}. 
Existing approaches, however, typically require a preconstructed determinant manifold whose optimal form depends on the physical regime; to ensure accuracy, one often includes a redundantly large set of Slater determinants before diagonalization. 
Here we introduce an active-learning extension of the adaptive truncated CI (ATCI) impurity solver~\cite{AraGo2017}. 
The Green's-function and self-energy formalism follows the ATCI framework, but the determinant space is constructed iteratively by active learning before the expensive Hamiltonian diagonalization step. 
Thus the present work adapts selection ideas familiar from quantum chemistry to the dynamical impurity problem, where Green's functions and self-energies, rather than only ground-state or low-lying excited-state energies, are the central quantities.

Within self-consistent DMFT, we find that the selected determinant manifold is controlled primarily by the correlation physics rather than by the nominal number of bath or auxiliary orbitals. 
Metallic systems require a broader selected space than insulating ones, reflecting stronger entanglement, but the approximation remains systematically controllable through the number of selected configurations. 
We demonstrate this behavior for the one-dimensional Hubbard model, reaching $N_c=10$ in cellular DMFT, corresponding to ten cluster sites in this single-orbital case,  and for Sr$_2$RuO$_4$, where dynamical quantities converge systematically as $N_b$ is increased from 9 to 18 in a three-orbital rotationally invariant impurity model. 
Measured in correlated impurity degrees of freedom, the present implementation reaches the ten-correlated-orbital scale, including one-band clusters with ten sites or multiorbital clusters with a comparable number of correlated orbitals.

% ==================================================================================================

% Method============================================================================================
\textit{Method.}---AL-ATCI builds on the adaptive-truncation configuration-interaction (ATCI) impurity solver~\cite{AraGo2017}, which evaluates ground-state properties, Green's functions, and self-energies in an iteratively optimized truncated determinant space.
In ATCI, candidate Slater determinants are generated from important reference configurations by particle-hole substitutions, and the Hamiltonian is constructed and diagonalized in the resulting expanded space.
The dominant cost is therefore the construction and diagonalization of a large many-body Hamiltonian before the irrelevant determinants have been eliminated.

AL-ATCI reduces this cost by inserting an active-learning selection step before Hamiltonian construction.
A random-forest classifier~\cite{Breiman2001,Pedregosa2011}, trained on data from previous CI iterations and DMFT cycles, ranks candidate Slater determinants by predicted importance~\cite{Coe2018,MejutoZaera2019,Jeong2021,Bilous2025}.
Only the top $N_{\text{query}}$ determinants are retained for Hamiltonian construction and diagonalization; $N_{\text{query}}$ is therefore the query size and the primary convergence parameter of AL-ATCI.

The machine-learning step is used only to select the determinant basis.
It does not modify the impurity Hamiltonian, the DMFT self-consistency condition, or the evaluation of observables.
Once the determinant set is selected, static quantities are obtained by ordinary diagonalization in that space, while Green's functions and self-energies are evaluated using the ATCI dynamical formalism~\cite{AraGo2017}.
Thus AL-ATCI inherits the real-frequency, zero-temperature impurity-solver structure of ATCI, while replacing exhaustive determinant expansion by a learned proposal step.
If all generated candidates are retained, the method reduces to ATCI; in the complete determinant-space limit it recovers ED for the finite impurity Hamiltonian.

The classifier ranks determinants, not orbitals.
Thus increasing $N_b$ enlarges the combinatorial space of possible determinants, but the number of physically important determinants grows much more weakly over the benchmark range studied here.
This is the mechanism behind the improved bath-size scalability.
Implementation details, including class-imbalance treatment, occupation encoding, symmetry handling, and dataset management, are given in the Supplemental Material~\cite{SM}.

%===================================================================================================

% Benchmark results=================================================================================
\textit{Benchmark on the one-dimensional Hubbard model.}---We benchmark AL-ATCI on the half-filled one-dimensional Hubbard model~\cite{Hubbard1963}:
\begin{align*}
\hat{H} = -t\sum_{\langle i,j \rangle,\sigma} \left(\hat{c}_{i\sigma}^{\dagger}\hat{c}_{j\sigma} + \text{H.c.}\right)
    - \mu \sum_{i,\sigma} \hat{n}_{i\sigma}
    + U \sum_{i} \hat{n}_{i\uparrow} \hat{n}_{i\downarrow},
\end{align*}
where $t$ is the nearest-neighbor hopping, $\mu = U/2$ enforces half-filling, and $U$ is the on-site repulsion.
This model is exactly solvable via the Bethe ansatz~\cite{Lieb1968,Penc1996,Capone2004,AraGo2009}, providing a stringent benchmark where mean-field approaches typically fail due to strong quantum fluctuations.
We employ cellular DMFT~\cite{Kotliar2001,Maier2005}, treating $N_c$ correlated impurity orbitals---cluster sites in this single-orbital model---coupled to $N_b$ bath orbitals.
All calculations are at zero temperature.
The solver benchmark throughout is against ED within the same DMFT setup; the Bethe-ansatz dispersion serves as a physical reference for spectral features.
Beyond this, the model provides a suitable testbed for independently studying the scaling with $N_c$ and $N_b$: since the cluster is a linear chain with nearest-neighbor hopping only, the point group symmetry of the cluster constrains the hybridization to couple strongly only to the endpoints.
This allows $N_c$ to be increased beyond $N_b$ once the bath discretization is sufficient~\cite{Koch2008}.
The benchmarks below separate three issues: convergence with $N_{\text{query}}$ at fixed $(N_c,N_b)$, scalability with $N_c$ at fixed $N_b$, and bath-size dependence with $N_b$ at fixed $N_c$.

% Systematic improvement based on the Nquery
\begin{figure}
    \centering
    \includegraphics[width=\columnwidth]{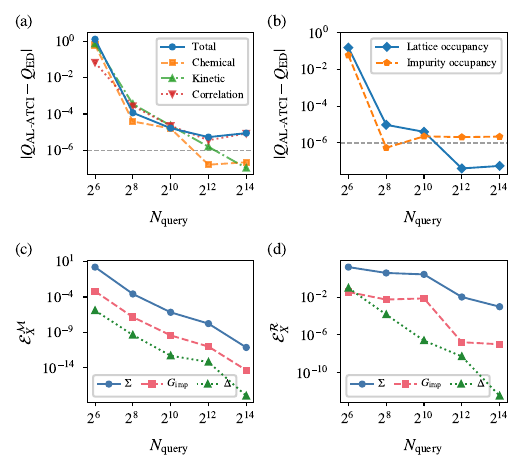}
    \caption{Convergence of AL-ATCI with query size $N_{\text{query}}$ for cellular DMFT
    with $(N_c, N_b) = (4, 8)$ at $U = 8t$ and zero temperature.
    (a) Absolute error in the total energy and its chemical-potential, kinetic, and correlation components relative to the ED reference.
    (b) Same for the lattice and impurity occupancies.
    (c), (d) Mean-squared Frobenius deviation
    $\mathcal{E}_X=\frac{1}{N_c^2 N_\omega}\sum_m \|\Delta\mathbf{X}(\omega_m)\|_F^2$
    for $X=\boldsymbol{\Sigma},\,G_{\text{imp}},\,\Delta$, evaluated on the Matsubara~(c) and real~(d) axes.%
    \label{fig:convergence}
}
\end{figure}

Figure~\ref{fig:convergence} demonstrates that the query size $N_{\text{query}}$ provides a direct and systematic control of accuracy.
Static quantities, such as the ground-state energy and lattice occupancy, converge smoothly toward the ED reference as $N_{\text{query}}$ increases, with even relatively small query sizes---corresponding to a tiny fraction of the full Hilbert space---already yielding near-quantitative agreement.
For dynamical quantities, we quantify the error by the mean-squared Frobenius deviation averaged over $N_\omega$ sampled frequency points on the corresponding Matsubara or real-frequency grid.
Figure~\ref{fig:convergence}(c,d) shows that the impurity Green’s function $G_{\rm imp}$, the self-energy $\Sigma$, and the hybridization function $\Delta$ all converge systematically on both the Matsubara and real-frequency axes.
Direct comparisons of $G_{\rm imp}$ with ED are provided in the Supplemental Material~\cite{SM}.
These results confirm that $N_{\text{query}}$ acts as a systematically improvable parameter controlling the accuracy of AL-ATCI across both static and dynamical observables.

% Scalability depend on (Nc, Nb)
\begin{figure}
    \centering
    \includegraphics[width=\columnwidth]{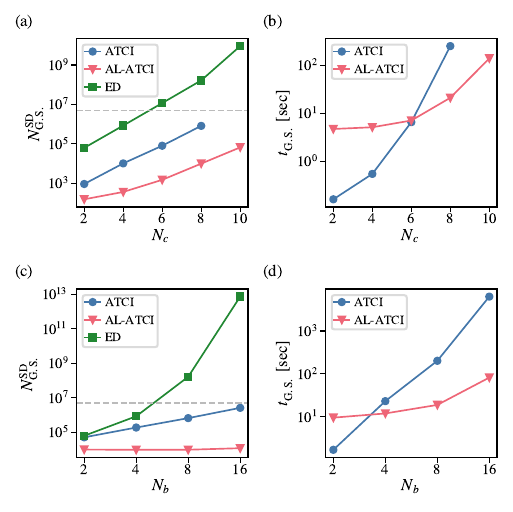}
    \caption{Computational scaling of AL-ATCI versus ATCI for the half-filled one-dimensional Hubbard model at $U/t = 8$ and zero temperature.
(a),~(b) Scaling with correlated impurity size $N_c$ (fixed $N_b = 8$):
number of selected Slater determinants in the ground-state calculation, $N^{\text{SD}}_{\text{G.S.}}$~(a),
and wall-clock time for the ground-state computation, $t_{\text{G.S.}}$~(b).
(c),~(d) Same quantities as a function of bath size $N_b$ (fixed $N_c = 8$).
All plotted quantities are averaged over the CI iterations.
Each data point corresponds to the minimum $N^{\text{SD}}_{\text{G.S.}}$
for which $\mathrm{Im}\,\Sigma(\omega) \leq 0$ is satisfied during the CI iterations, used here as an operational internal convergence marker
(see Supplemental Material~\cite{SM}).
The dashed line marks the practical diagonalization limit
($N^{\text{SD}}_{\text{G.S.}} = 5\times10^6$).
\label{fig:scalability}}
\end{figure}

Figure~\ref{fig:scalability} compares the scaling of AL-ATCI and ATCI with correlated impurity size $N_c$ and bath size $N_b$.
For the scaling analysis, we use the first restoration of causality, $\mathrm{Im}\,\Sigma(\omega)\le 0$, as an operational internal convergence marker and report the corresponding configuration count and wall-clock time.
This criterion is not used as a stand-alone claim of exactness; rather, it identifies the onset of a physically acceptable solution, after which explicit $N_{\text{query}}$ scans provide systematic refinement and, where ED is available, convergence to indistinguishable self-energies.
As $N_c$ increases [Fig.~\ref{fig:scalability}(a, b)], both the configuration count and the computational cost grow rapidly, but significantly more so for ATCI than for AL-ATCI, with the latter becoming advantageous beyond $N_c \approx 6$.

A more pronounced difference emerges in the bath-size scaling [Fig.~\ref{fig:scalability}(c, d)].
While ATCI exhibits a steep increase in both configuration count and wall-clock time, AL-ATCI shows a markedly reduced dependence on $N_b$, with a substantially weaker growth of computational cost.
This behavior reflects the fact that, although the total number of possible Slater determinants increases rapidly with bath size, only a relatively small subset contributes significantly to the ground-state wavefunction, which AL-ATCI identifies directly.

This improved bath scaling has important practical consequences.
It enables the use of larger bath sizes, thereby reducing discretization errors in representing the hybridization function, and allows for a more systematic assessment of bath convergence.
Moreover, at the level of the determinant-selection bottleneck addressed here, increasing $N_b$ and appending additional noninteracting orbitals relevant to finite-Hamiltonian XAS/RIXS calculations pose the same combinatorial challenge, so enlarged-orbital calculations incur comparatively little additional cost.

% Improved momentum resolution on spectral function
\begin{figure}
    \centering
    \includegraphics[width=\columnwidth]{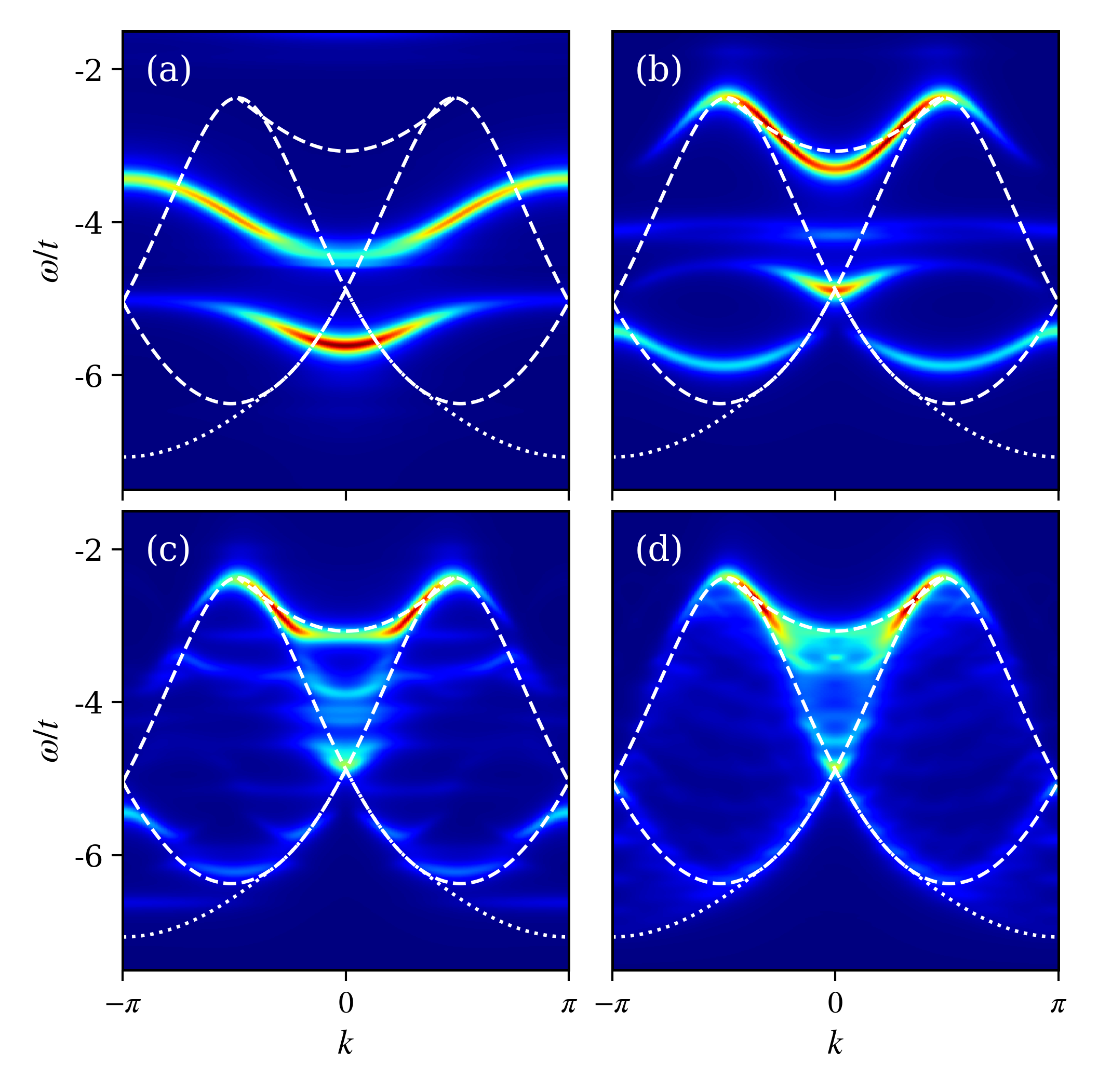}
    \caption{Spectral function from cellular DMFT for the one-dimensional Hubbard model at $U=8t$, zero temperature, and fixed total bath size $N_b=8$.
    (a) $N_c = 1$ (single-site DMFT), (b) $N_c = 2$, (c) $N_c = 4$, (d) $N_c = 10$.
    Larger clusters capture richer short-range correlations within the cellular DMFT approximation.
    White dashed (dotted) lines: spinon (holon) dispersion from the Bethe ansatz~\cite{Lieb1968}.}
    \label{fig:spectral_evolution}
\end{figure}

Figure~\ref{fig:spectral_evolution} shows the effect of increasing cluster size at fixed total bath size $N_b=8$.
The purpose is not to demonstrate bath convergence, but to isolate the gain from cluster enlargement once the bath already captures the essential hybridization physics~\cite{Koch2008}.
Even the change from $N_c=1$ to $N_c=2$ qualitatively reshapes the spectrum, while larger clusters bring the dominant features closer to the Bethe-ansatz spinon and holon branches.
The $N_c=10$ result illustrates the cluster-DMFT regime made practical by AL-ATCI.
%===================================================================================================

% t2g model=========================================================================================
\begin{figure}
    \centering
    \includegraphics[width=\columnwidth]{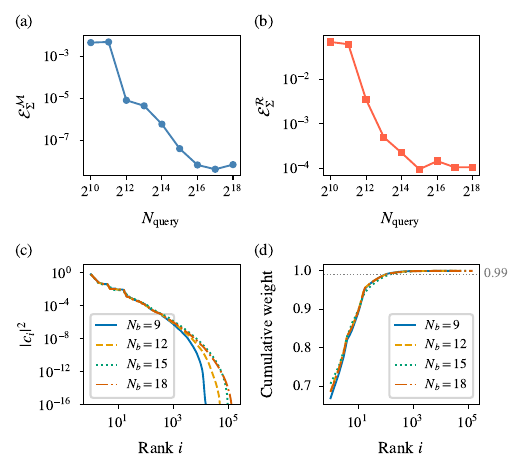}
    \caption{Convergence of AL-ATCI for the multi-orbital impurity problem.
        (a) Mean-squared Frobenius deviation of the self-energy,
        $\mathcal{E}_{\Sigma}=\frac{1}{N_c^2 N_\omega}\sum_m \|\Delta\boldsymbol{\Sigma}(\omega_m)\|_F^2$,
        as a function of query size $N_{\text{query}}$, evaluated on the Matsubara axis.
        (b) Same as (a) but evaluated on the real axis.
        (c) CI coefficient weights $|c_i|^2$ ranked in descending order for bath sizes $N_b = 9, 12, 15, 18$,
        showing rapid decay that justifies sparse truncation of the Hilbert space.
        (d) Cumulative weight $\sum_{j \leq i} |c_j|^2$ as a function of rank $i$ for the same $N_b$ values.
        \label{fig:SRO}}
\end{figure}

\textit{Application to a multiorbital system.}—We now test AL-ATCI on Sr$_2$RuO$_4$, a prototypical three-orbital $t_{2g}$ Hund metal~\cite{Mravlje2011,Georges2013}.
This serves as a demanding multiorbital impurity benchmark: the solver must handle the full rotationally invariant Slater–Kanamori interaction~\cite{Kanamori1963} (see Supplemental Material~\cite{SM} for the explicit Hamiltonian), and the three-orbital structure produces a substantially larger Hilbert space than the single-band case.
We use a three-band tight-binding model with hopping parameters and interaction values ($U=2.5$\,eV, $J=0.4$\,eV) from Ref.\cite{AraGo2020}, fixing the total occupancy at four electrons per unit cell ($d^4$ filling).
Spin–orbit coupling is neglected; since Sr$_2$RuO$_4$ is known to be sensitive to spin–orbit coupling~\cite{Haverkort2008}, the present calculation should be viewed as a solver benchmark rather than a complete materials description.
Bath parameters are determined by weighted least-squares fitting on the Matsubara axis ($1/\omega_n$ weighting).

Figure~\ref{fig:SRO} demonstrates systematic convergence of the self-energy with increasing query size, both on the Matsubara and real-frequency axes~(a, b), confirming that the accuracy of AL-ATCI remains controllable in the multiorbital setting.
In addition, the distribution of CI coefficients exhibits a rapid decay across all bath sizes considered~(c), with the cumulative weight reaching near saturation with only a small fraction of configurations~(d).
This behavior provides direct evidence that the many-body wavefunction remains highly compressible even in the enlarged Hilbert space of the three-orbital problem, thereby justifying the efficiency of the truncated configuration space.
At the level of the Hilbert-space bottleneck addressed here, increasing $N_b$ and appending additional noninteracting orbitals pose the same determinant-selection problem.
The large-$N_b$ multiorbital benchmark therefore directly probes the mechanism that makes enlarged-orbital finite-Hamiltonian calculations more practical.

The key point is that this convergence is achieved in a fully rotationally invariant three-orbital problem, where conventional ED is unavailable at the target bath sizes.
Convergence is instead controlled internally through $N_{\text{query}}$, while the compressed CI-weight distributions across $N_b=9$–$18$ show that the determinant-space compression underlying the weak bath-size dependence of AL-ATCI persists in a realistic multiorbital impurity problem.
%===================================================================================================

% conclusions=======================================================================================
\textit{Conclusions.}---We have introduced AL-ATCI, a finite-Hamiltonian impurity solver that combines the ATCI Green's-function and self-energy formalism with active-learning selection of the determinant manifold.
The central result is that, over the benchmark range studied here, the relevant computational cost is controlled by the physically important Slater determinants rather than by the full combinatorial Hilbert space.
The query size $N_{\rm query}$ therefore provides a systematically improvable convergence knob, even when conventional ED is unavailable.

The benchmarks demonstrate two complementary gains.
At fixed correlated impurity size, AL-ATCI enables larger bath discretizations, as shown by the rotationally invariant three-orbital Sr$_2$RuO$_4$ problem up to $N_b=18$.
At fixed total bath size, it makes cluster enlargement practical, reproducing ED-level results for the one-dimensional Hubbard model and reaching cellular-DMFT clusters up to $N_c=10$, where $N_c$ counts correlated impurity orbitals.
The same mechanism also supports enlarged finite-Hamiltonian impurity models for core-level spectroscopies: adding ligand or core orbitals for XAS/RIXS raises the same determinant-selection bottleneck as adding bath orbitals, so the weak dependence on added one-particle orbitals found here should make such calculations substantially more practical.

This capability is especially relevant for low-dimensional multiorbital materials, where short-range spatial correlations are essential.
For example, $\alpha$-RuCl$_3$ is a quasi-two-dimensional honeycomb Kitaev quantum-spin-liquid candidate~\cite{Plumb2014,Kim2015,Sandilands2015,Banerjee2016,Kasahara2018}.
A two-site $t_{2g}$ cluster contains six correlated orbitals and already captures bond-dependent correlations absent in single-site DMFT, placing it well within the correlated-orbital scale demonstrated here.
The remaining limitation is that the required selected space remains problem dependent: metallic or highly entangled systems require larger $N_{\rm query}$.
Within this limitation, AL-ATCI relaxes the bath-discretization bottleneck of ED/CI impurity solvers while preserving their direct real-frequency access and explicit-Hamiltonian flexibility, opening a practical route to large-bath, multiorbital cluster, and enlarged-orbital impurity calculations.%===================================================================================================

\begin{acknowledgments}
The authors thank Young-Woo Son, Woo-Seok Jeong, and Andrew J. Millis for helpful discussions.
This work was supported by the National Research Foundation of Korea (NRF) under Grants No. NRF-2021R1C1C1010429, NRF-2023M3K5A1094813, RS-2023-00256050, RS-2024-00354199, and RS-2025-00515360.
\end{acknowledgments}

\bibliographystyle{apsrev4-2}
\bibliography{main_ref}

%==================== Supplemental Material ====================
\clearpage
\onecolumngrid
\appendix

\setcounter{figure}{0}
\setcounter{table}{0}
\setcounter{equation}{0}
\setcounter{section}{0}

\renewcommand{\thefigure}{S\arabic{figure}}
\renewcommand{\thetable}{S\arabic{table}}
\renewcommand{\theequation}{S\arabic{equation}}

\section*{Supplemental Material}
\noindent
\begin{center}
    \textbf{A Scalable Configuration-Interaction Impurity Solver via Active Learning}
\end{center}

\section{Implementation Details of AL-ATCI}
Each configuration is represented as a binary vector encoding the spin-orbital occupations of all single-particle states.
The training labels are assigned based on the magnitude of the wavefunction coefficient: configurations whose absolute coefficient exceeds a threshold are labeled as important, and all remaining configurations as unimportant; in the present work, we use a threshold of $10^{-6}$.
Because important configurations constitute a small fraction of the total candidate space, the training set is severely class-imbalanced; we address this by assigning class weights inversely proportional to the number of samples in each class.

When the top $N_{\text{query}}$ configurations are selected, their time-reversal partners are appended to preserve time-reversal symmetry.
As a result, the total number of Slater determinants $N_{\text{SD}}^{\text{G.S.}}$ in the ground-state wavefunction can exceed $N_{\text{query}}$; the ratio $N_{\text{SD}}^{\text{G.S.}}/N_{\text{query}}$ does not substantially exceed unity in practice, as shown in Table~\ref{nSD}.

The training dataset accumulates across AL-ATCI iterations within a single DMFT cycle and is reset at the beginning of each new DMFT cycle.
This reset is necessary because updated bath parameters alter the hybridization structure, rendering the configuration importance data from the previous cycle unrepresentative of the new impurity problem.

\begin{figure}
    \centering
    \includegraphics[width=0.5\columnwidth]{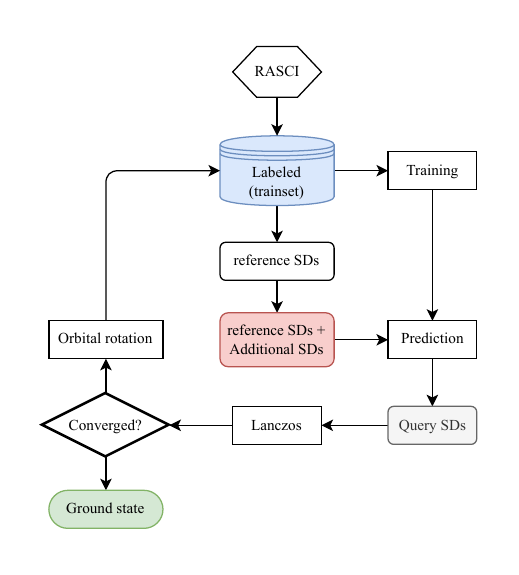}
    \caption{Flowchart of AL-ATCI, highlighting the active learning enhancements: classifier training, configuration importance evaluation, and selective querying of top-ranked configurations.}
    \label{fig:flowchart}
\end{figure}

\clearpage

\begin{figure}[tbph]
\centering
\includegraphics[width=\textwidth]{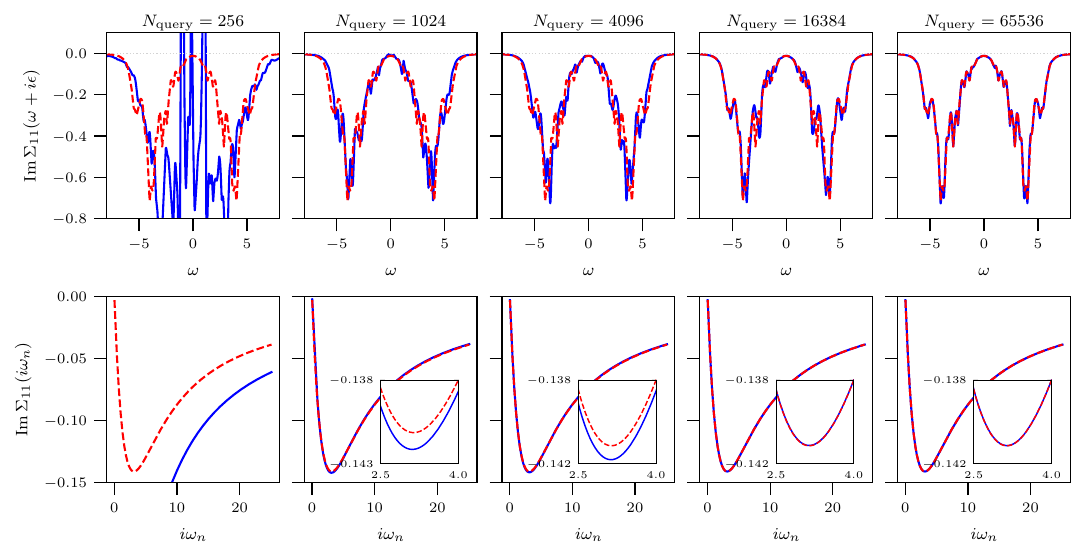}
\caption{Convergence of the impurity self-energy with increasing $N_{\text{query}}$ at $U/t = 2$.
Top row: $\text{Im}\,\Sigma(\omega)$ on the real-frequency axis.
Bottom row: $\text{Im}\,\Sigma(i\omega_n)$ on the Matsubara-frequency axis, shown for 512 Matsubara frequencies at $\beta t = 128$.
AL-ATCI results for $N_{\text{query}} = 256, 1024, 4096, 16384$, and $65536$ (blue lines) are compared against the ED benchmark (red dashed lines).
Parameters: $N_c = 4$, $N_b = 8$, half-filling, broadening $\epsilon = 0.1t$.}
\label{fig:selfenergy_u2.0}
\end{figure}

\section{Self-Energy Convergence with Query Size $N_{\text{query}}$}
\begin{figure}[htbp]
\centering
\includegraphics[width=\textwidth]{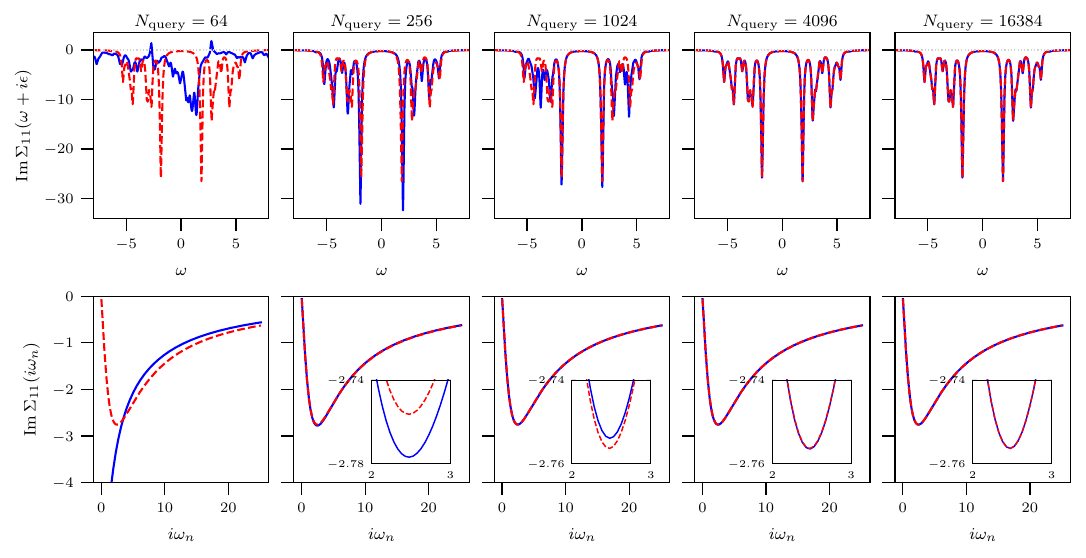}
\caption{Convergence of the impurity self-energy with increasing $N_{\text{query}}$ at $U/t = 8$.
Top row: $\text{Im}\,\Sigma(\omega)$ on the real-frequency axis.
Bottom row: $\text{Im}\,\Sigma(i\omega_n)$ on the Matsubara-frequency axis, shown for 512 Matsubara frequencies at $\beta t = 128$.
AL-ATCI results for $N_{\text{query}} = 64, 256, 1024, 4096$, and $16384$ (blue lines) are compared against the ED benchmark (red dashed lines).
At $N_{\text{query}} = 64$, the real-frequency self-energy exhibits a causality violation, which is resolved from $N_{\text{query}} = 256$ onward.
Parameters: $N_c = 4$, $N_b = 8$, half-filling, broadening $\epsilon = 0.1t$.}
\label{fig:selfenergy_u8.0}
\end{figure}

In the main text, Fig.~1 demonstrates that various physical quantities computed with AL-ATCI converge to exact diagonalization (ED) results as the number of active learning queries $N_{\text{query}}$ increases.
Here we extend this analysis to the impurity self-energy, shown on both the real- and Matsubara-frequency axes as $\Sigma(\omega+i\epsilon)$ and $\Sigma(i\omega_n)$.
Figure~\ref{fig:selfenergy_u2.0} shows the imaginary part of the self-energy on both the real and Matsubara frequency axes for $N_c = 4$, $N_b = 8$, at $U/t = 2$, with columns corresponding to $N_{\text{query}} = 256, 1024, 4096, 16384$, and $65536$.
Figure~\ref{fig:selfenergy_u8.0} shows the same quantities at $U/t = 8$, with columns corresponding to $N_{\text{query}} = 64, 256, 1024, 4096$, and $16384$.
In both cases, the AL-ATCI results (blue lines) systematically converge toward the ED benchmark (red dashed lines) as $N_{\text{query}}$ increases, and the Matsubara-frequency self-energy converges more rapidly than its real-frequency counterpart.

At $U/t = 2$, the result at $N_{\text{query}} = 256$ exhibits a violation of causality, which is restored from $N_{\text{query}} = 1024$ onward.
The AL-ATCI and ED self-energies become visually indistinguishable at $N_{\text{query}} = 65536$, requiring a larger $N_{\text{query}}$ than the $U/t = 8$ case.

At $U/t = 8$, the ground state CI expansion is sparse, with a small number of configurations carrying most of the weight, making it particularly amenable to the adaptive truncation strategy of AL-ATCI.
The result at $N_{\text{query}} = 64$ violates causality, and although causality is restored at $N_{\text{query}} = 256$, quantitative discrepancies with the ED benchmark persist at intermediate frequencies.
These discrepancies are systematically reduced with increasing $N_{\text{query}}$, and the AL-ATCI and ED self-energies become visually indistinguishable at $N_{\text{query}} = 4096$.
Notably, this convergence is achieved at a smaller $N_{\text{query}}$ than the $U/t = 2$ case, suggesting that AL-ATCI is particularly efficient for insulating systems where the ground state is dominated by a smaller number of important configurations.

These results confirm that AL-ATCI provides systematically improvable approximations to the self-energy across both the metallic and insulating regimes.
The convergence rate reflects the underlying correlation strength and ground-state structure, with insulating systems requiring fewer configurations to reach convergence than their metallic counterparts.

\begin{figure}[htbp]
\centering
\includegraphics[width=\textwidth]{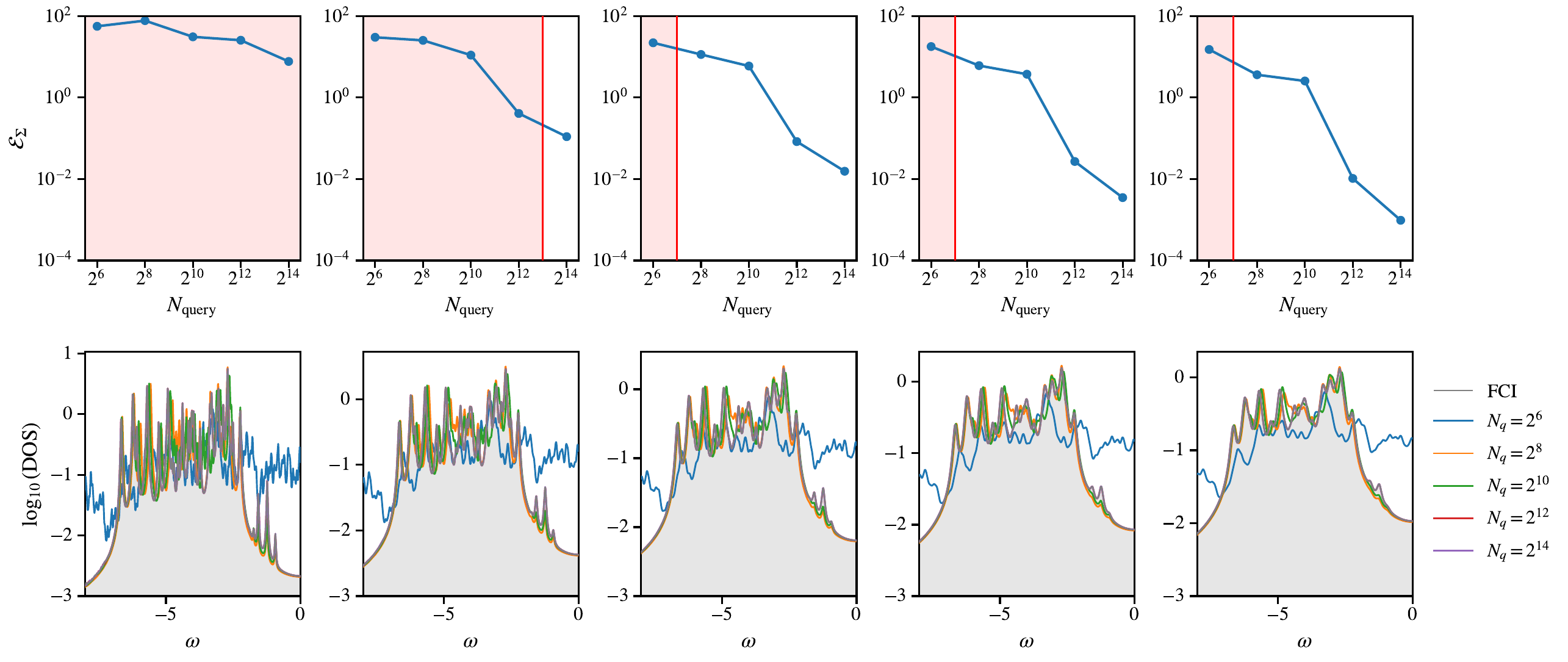}
\caption{Convergence of AL-ATCI with increasing $N_{\text{query}}$ at $U/t = 8$ for different broadening values.
Each column corresponds to a distinct broadening $\epsilon = 0.02, 0.04, 0.06, 0.08, 0.1$ (in units of $t$) from left to right.
Top row: mean squared Frobenius norm of the real-frequency self-energy deviation
$\mathcal{E}_{\Sigma}=\frac{1}{N_c^2 N_\omega}\sum_m \|\Delta\boldsymbol{\Sigma}(\omega_m + i\epsilon)\|_F^2$ as a function of $N_{\text{query}}$.
The red vertical line marks the value of $N_{\text{query}}$ at which the causality condition $\mathrm{Im}\,\Sigma(\omega) \leq 0$ is first satisfied.
Bottom row: $\log_{10}(\mathrm{DOS})$ for ED (grey) and AL-ATCI with $N_{\text{query}} = 64$ (blue), $256$ (orange), $1024$ (green), $4096$ (red), and $16384$ (purple).
Parameters: $N_c = 4$, $N_b = 8$, half-filling.}
\label{fig:convergence_Nq}
\end{figure}

Figure~\ref{fig:convergence_Nq} further illustrates this convergence for varying broadening $\epsilon$.
The top row shows $\mathcal{E}_{\Sigma}$ as a function of $N_{\text{query}}$ for five broadening values $\epsilon = 0.02$--$0.1t$: in all cases the deviation decreases monotonically, and the onset of causality (red vertical line) occurs at a consistent $N_{\text{query}}$, confirming that the convergence behavior is robust across broadening values.
The bottom row shows $\log_{10}(\mathrm{DOS})$ at the lower Hubbard band, where the spectral features converge systematically toward the ED reference with increasing $N_{\text{query}}$.
Note that the two smallest broadening values ($\epsilon = 0.02t$ and $0.04t$) are sufficiently small that even the ED reference can exhibit causality violations; for the remaining values $\epsilon = 0.06$--$0.1t$, the $N_{\text{query}}$ at which causality is first restored (red vertical line) is consistent across broadening values, confirming that the convergence behavior of AL-ATCI is largely insensitive to the choice of $\epsilon$.

\begin{table}[htbp]
\centering
\begin{tabular}{rr|rr}
\hline
\multicolumn{2}{c|}{$U/t=2$} & \multicolumn{2}{c}{$U/t=8$} \\
\hline
$N_{\text{query}}$ & $N_{\text{SD}}^{\text{G.S.}}$ & $N_{\text{query}}$ & $N_{\text{SD}}^{\text{G.S.}}$ \\
\hline
$256$      & $268$      & $64$       & $84$ \\
$1\,024$   & $1\,242$   & $256$      & $284$ \\
$4\,096$   & $4\,350$   & $1\,024$   & $1\,186$ \\
$16\,384$  & $18\,574$  & $4\,096$   & $4\,469$ \\
$65\,536$  & $74\,518$  & $16\,384$  & $18\,866$ \\
\hline
\end{tabular}
\caption{Number of Slater determinants contributing to the ground state.
$N_{\text{query}}$ denotes the number of queried configurations in the selected-CI procedure, while $N_{\text{SD}}^{\text{G.S.}}$ is the number of Slater determinants with nonzero weight in the resulting ground-state wavefunction.
Because time-reversal symmetry is enforced, each queried configuration generates its time-reversed partner, so that $N_{\text{SD}}^{\text{G.S.}}$ can be larger than $N_{\text{query}}$.
Results are shown for $U/t=2$ and $U/t=8$.}
\label{nSD}
\end{table}

\clearpage

\section{Details of Scalability Measurement}
\begin{figure}[htbp]
\centering
\includegraphics[width=0.6\textwidth]{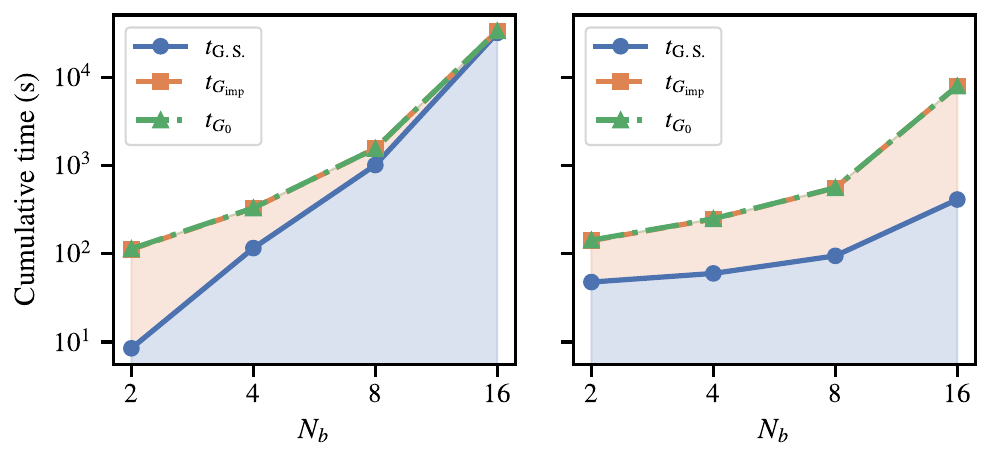}
\caption{Cumulative wall-clock time as a function of bath size $N_b$ for ATCI (left) and AL-ATCI (right).
Three timing components are shown: ground-state diagonalization $t_{\text{G.S.}}$, impurity Green's function evaluation $t_{G_{\text{imp}}}$, and bath Green's function evaluation $t_{G_0}$.
The shaded region highlights the gap between $t_{\text{G.S.}}$ and the Green's function timings, indicating the relative cost of diagonalization.
Parameters: $N_c = 8$, half-filling, $U/t = 8$, zero temperature.}
\label{fig:timing_nb}
\end{figure}

\begin{figure}[htbp]
\centering
\includegraphics[width=0.6\textwidth]{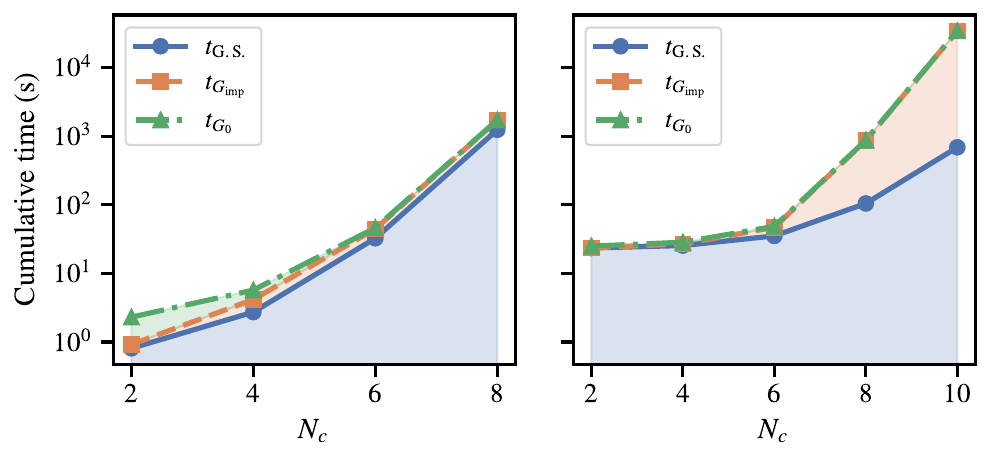}
\caption{Cumulative wall-clock time as a function of cluster size $N_c$ for ATCI (left) and AL-ATCI (right).
Three timing components are shown: ground-state diagonalization $t_{\text{G.S.}}$, impurity Green's function evaluation $t_{G_{\text{imp}}}$, and bath Green's function evaluation $t_{G_0}$.
The shaded region highlights the gap between $t_{\text{G.S.}}$ and the Green's function timings, indicating the relative cost of diagonalization.
Parameters: $N_b = 8$, half-filling, $U/t = 8$, zero temperature.}
\label{fig:timing_nc}
\end{figure}

Each data point in Fig.~2 of the main text corresponds to the minimum number of configurations $N_{\text{SD}}^{\text{G.S.}}$ for which the resulting self-energy satisfies the causality condition $\mathrm{Im}\,\Sigma(\omega) \leq 0$ for all frequencies, together with the corresponding wall-clock time.
All timing measurements were performed on a single node equipped with two Intel Xeon Platinum 8360Y processors (36 cores each, 72 cores in total, 2.40\,GHz base clock, 48\,KiB L1d and 32\,KiB L1i per core, 1.25\,MiB L2 per core, 54\,MiB L3 per socket), using OpenMP parallelization across all available cores.
Timing measurements are restricted to the first five CI iterations to ensure a consistent comparison across system sizes.
This restriction is necessary because the selected configuration space can evolve differently in later iterations as the natural-orbital basis and the training dataset both evolve, making a direct comparison between AL-ATCI and ATCI ambiguous beyond the early iterations.
Figures~\ref{fig:timing_nb} and~\ref{fig:timing_nc} show the breakdown of wall-clock time into ground-state diagonalization $t_{\text{G.S.}}$, impurity Green's function evaluation $t_{G_{\text{imp}}}$, and bare Green's function evaluation $t_{G_0}$, as a function of $N_b$ and $N_c$, respectively, confirming that diagonalization dominates the total cost in both ATCI and AL-ATCI.
The remaining contributions---particle-hole substitution enumeration, Hamiltonian construction, and the machine-learning train/predict overhead---are too small to be visible on the scale of the figures and are therefore not shown separately.

\clearpage

\section{Momentum-Resolved Spectral Function Convergence}
Here we examine how the momentum-resolved spectral function $A(k,\omega)$ evolves with increasing cluster size, complementing Fig.~3 of the main text.
To demonstrate the systematic convergence of AL-ATCI with increasing cluster size, we compute momentum-resolved spectral functions $A(k, \omega)$ for various cluster sizes at representative momentum points in the Brillouin zone.

\begin{figure}[htbp]
\centering
\includegraphics[width=\textwidth]{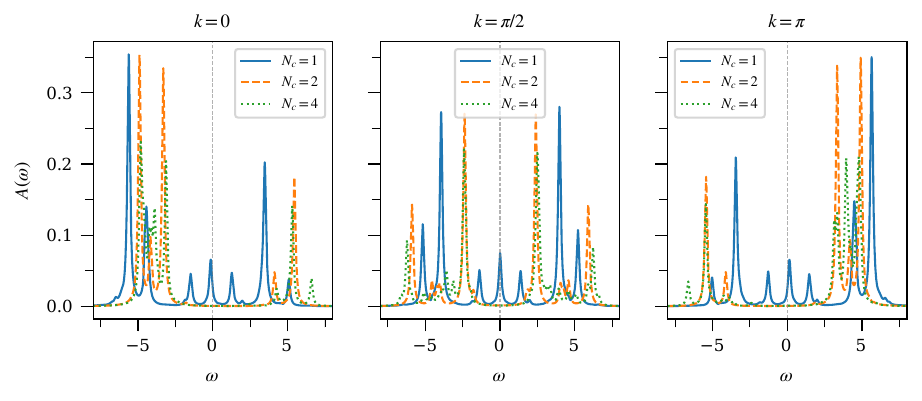}
\caption{Momentum-resolved spectral functions $A(k, \omega)$ for cluster sizes $N_c = 1$ (blue), $N_c = 2$ (orange), and $N_c = 4$ (green) at three representative momentum points.
From left to right: $k = 0$, $\pi/2$, $\pi$.
Parameters: $U/t = 8$, $N_b = 8$, half-filling, broadening $\epsilon = 0.1t$.}
\label{fig:spectral_small}
\end{figure}

Figure~\ref{fig:spectral_small} shows the spectral functions at three momentum points ($k = 0, \pi/2, \pi$) for cluster sizes $N_c = 1, 2, 4$.
As $N_c$ increases, the spectral features evolve systematically, with peak positions, heights, and spectral weights showing clear convergence with cluster size.
In particular, the spurious midgap spectral weight present at $N_c = 1$ vanishes for larger $N_c$, consistent with the Mott-insulating phase at $U/t = 8$.
The particle-hole symmetry $A(k,\omega) = A(\pi-k,-\omega)$ is clearly satisfied: the spectra at $k=0$ and $k=\pi$ are mirror images of each other under $\omega \rightarrow -\omega$, while the spectrum at $k=\pi/2$ is symmetric about $\omega=0$.

\begin{figure}[htbp]
\centering
\includegraphics[width=\textwidth]{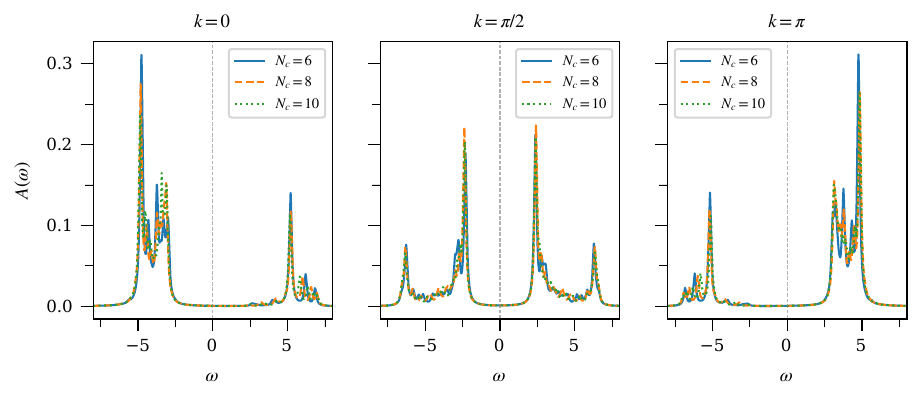}
\caption{Momentum-resolved spectral functions $A(k, \omega)$ for cluster sizes $N_c = 6$ (blue), $N_c = 8$ (orange), and $N_c = 10$ (green) at three representative momentum points.
From left to right: $k = 0$, $\pi/2$, $\pi$.
Parameters: $U/t = 8$, $N_b = 8$, half-filling, broadening $\epsilon = 0.1t$.}
\label{fig:spectral_large}
\end{figure}

Figure~\ref{fig:spectral_large} extends this analysis to larger cluster sizes $N_c = 6, 8, 10$, which are inaccessible to conventional ED due to the exponential growth of the Hilbert space.
The spectral functions at $N_c=8$ and $N_c=10$ differ only weakly from those at $N_c=6$, indicating that the spectra are already close to convergence by $N_c=6$.
The particle-hole symmetry $A(k,\omega) = A(\pi-k,-\omega)$ is again clearly preserved across all cluster sizes.

These results confirm that AL-ATCI achieves systematic convergence of the momentum-resolved spectral function with increasing $N_c$, correctly reproducing the Mott gap and the particle-hole symmetry of the half-filled Hubbard model at $U/t = 8$.
The ability to reach cluster sizes well beyond the reach of conventional ED demonstrates the practical advantage of the active learning approach in accessing the large-$N_c$ regime.

\clearpage

\section{Slater-Kanamori Hamiltonian and Spectral Function of $\text{Sr}_2\text{RuO}_4$}
\begin{figure}[h]
\centering
\includegraphics[width=\textwidth]{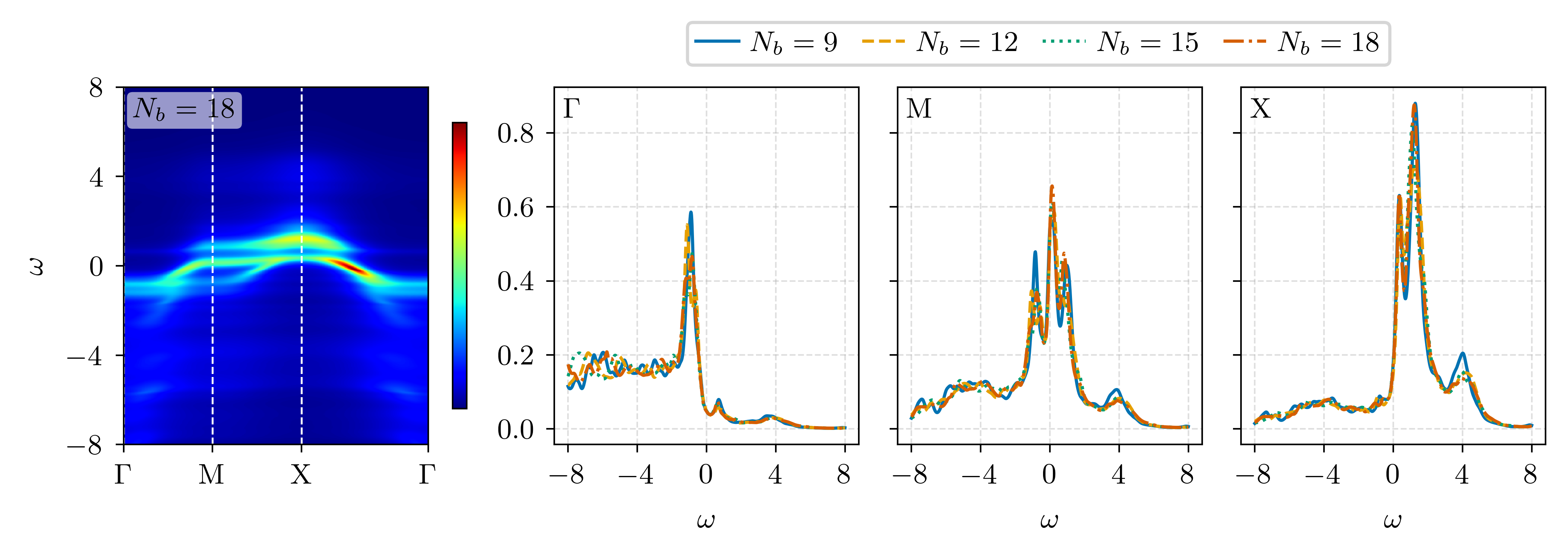}
\caption{Spectral function of Sr$_2$RuO$_4$ from zero-temperature single-site DMFT with AL-ATCI, for $U = 2.5$\,eV and $J = 0.4$\,eV.
As $N_b$ increases from 9 to 18, the spectral function converges systematically, confirming that AL-ATCI's weak bath-size dependence extends to a three-orbital problem with the full rotationally invariant interaction.}
\label{fig:SRO_SM}
\end{figure}

In the Sr$_2$RuO$_4$ calculations presented in the main text, the local interaction is described by the rotationally invariant Slater-Kanamori Hamiltonian for the $t_{2g}$ manifold,
\begin{align}
H_{\text{int}} &= U \sum_{m} n_{m\uparrow} n_{m\downarrow}
+ U' \sum_{m < m'} \sum_{\sigma \sigma'} n_{m\sigma} n_{m'\sigma'} \notag \\
&\quad - J \sum_{m < m'} \sum_{\sigma \sigma'} c^{\dagger}_{m\sigma} c_{m\sigma'} c^{\dagger}_{m'\sigma'} c_{m'\sigma} \notag \\
&\quad + J \sum_{m < m'} \left( c^{\dagger}_{m\uparrow} c^{\dagger}_{m\downarrow} c_{m'\downarrow} c_{m'\uparrow} + \text{H.c.} \right),
\end{align}
where $m, m' \in \{d_{xy}, d_{xz}, d_{yz}\}$ label the $t_{2g}$ orbitals, $\sigma$ denotes spin, and $n_{m\sigma} = c^{\dagger}_{m\sigma} c_{m\sigma}$.
The parameters $U$, $U' = U - 2J$, and $J$ denote the intraorbital Coulomb repulsion, interorbital Coulomb repulsion, and Hund's coupling, respectively.
The relation $U' = U - 2J$ ensures rotational invariance in orbital space.
In the main text, we use $U = 2.5$\,eV and $J = 0.4$\,eV.
Figure~\ref{fig:SRO_SM} shows the zero-temperature spectral function for $N_b = 9, 12, 15, 18$.
Both $xy$ and $xz/yz$ components converge systematically with increasing bath size, displaying well-defined quasiparticle bands near the Fermi level and incoherent weight at higher energies, consistent with the correlated metallic state of Sr$_2$RuO$_4$.
This confirms that the weak bath-size dependence observed in the single-band case extends to the three-orbital rotationally invariant problem.

%===============================================================

\end{document}